# Méthodologie de modélisation et d'implémentation d'adaptateurs spatio-temporels

**C. Chavet[‡†], P. Coussy[‡], P. Urard[†], E. Martin[‡]**

Laboratoire LESTER[‡] – Université de Bretagne Sud (France), CNRS FRE 2734
Central R&D[†] – STMicroelectronics, Crolles (France)
E-mail: prénom.nom@univ-ubs.fr ou prénom.nom@st.com

**Résumé :** La réutilisation de blocs préconçus est un concept connu du développement logiciel. Cette technique s'applique depuis quelques années aux systèmes sur silicium (SoC) dont la complexité et l'hétérogénéité sont grandissantes. La réutilisation se fait au niveau composants, appelés composants virtuels (IP), disponibles sous des formes plus ou moins flexibles. Ces composants sont des blocs fonctionnels dédiés : au traitement du signal (DCT, FFT), aux télécommunications (Viterbi, TurboCodes), … Ces blocs fonctionnels reposent sur un modèle d'architecture figée avec des degrés de personnalisation très réduits. Cette rigidité est particulièrement vraie pour l'interface de communication dont les ordres d'acquisition et de production de données, le comportement temporel et les protocoles d'échanges sont figés. L'intégration réussie d'un IP requière que le concepteur (1) synchronise les composants (2) convertisse les protocoles entre blocs "incompatibles" (3) temporise les données pour garantir les contraintes temporelles et l'ordre des données. Cette phase reste cependant très manuelle et source d'erreurs. Notre approche propose une modélisation formelle, à base de graphes de transferts de données, des flots de données entre composants. Le flot de synthèse s'appuie sur un ensemble de transformations du graphe initial pour aboutir à une architecture d'interface permettant l'adaptation spatio-temporelle des échanges de données entre plusieurs composants.

**Mots Clés :** Modélisation formelle, SoC, interface, systèmes embarqués, synthèse de haut niveau.

## 1 INTRODUCTION

### 1.1 Généralités

Les applications du traitement du signal (TDSI) sont maintenant largement utilisées dans des domaines variés allant de l'automobile aux communications sans fils, en passant par les applications multimédias et les télécommunications. La complexité croissante des algorithmes implémentés, et l'augmentation continue des volumes de données et des débits applicatifs, requièrent souvent la conception de circuits intégrés dédiés (ASIC). Typiquement l'architecture d'un composant complexe du TDSI utilise (1) des éléments de calculs de plus en plus complexes, (2) des mémoires et des modules de brassage de données (entrelaceur/désentrelaceur pour les TurboCodes, blocs de redondance spatio-temporelle dans les systèmes OFDM[1]/MIMO, …) et (3) privilégie des connexions point à point pour la communication inter éléments de calcul. Aujourd'hui, le coût de ces systèmes en terme d'élément mémorisant est très élevé; les concepteurs cherchent donc à minimiser la taille de ces buffers afin de réduire la consommation et la surface total du circuit, tout en cherchant à en optimiser les performances. Sur cette problématique globale, nous nous intéressons à l'optimisation des interfaces de communication entre composants. On peut voir ce problème comme la synthèse (1) des interfaces des IP, (2) de composant de brassage de données à part entière (type entrelaceur).

Pour la conception de la partie calculatoire du système, les concepteurs peuvent (1) réutiliser des blocs IP préconçus, et/ou (2) utiliser des outils de synthèse de haut niveau afin d'optimiser le processus de traduction d'un algorithme en un circuit intégré. Malheureusement, l'hypothèse selon laquelle les IP peuvent être utilisés sans modifications pour la conception de différents systèmes s'applique bien aux composants de type processeur (DSP, …) mais n'est pas réaliste pour des coprocesseurs de type TDSI (FFT, …). La synchronisation et la communication de données entre composants réutilisés requièrent souvent la conception d'une interface supplémentaire car ces composants n'ont pas été pensés et conçus pour communiquer et échanger des données entre eux. Ainsi, chaque combinaison de composants réutilisés implique la conception d'une interface différente, ce qui devient rapidement une tâche lourde dans le flot de conception d'un système (adaptation des contraintes temporelles, des séquences de transferts, de la synchronisation …). Une solution peut être d'utiliser un outil de synthèse de haut niveau pour générer des architectures à partir d'une description algorithmique du traitement à réaliser.

Pour des composants de type entrelaceur (brassage de données), typiquement utilisés dans les applications à base de TurboCodes, la rapidité d'évolution des standards, et la difficulté d'une conception manuelle, en font une problématique majeure pour la conception de circuits. De fait, leur conception peut amener à des solutions coûteuses en latence, en mémoires et en consommation. Une difficulté majeure vient notamment du fait que ces composants peuvent avoir différents modes de fonctionnement, et passer de l'un à l'autre en cours d'exécution. Les concepteurs doivent alors non seulement optimiser les architectures de chacun de ces modes, mais il leur faut également fusionner ces différentes configurations en un seul et même circuit, sans pour autant engendrer un surcoût trop important (en se contentant d'accoler les différents chemins de données par exemple), ce qui suppose de trouver l'architecture la plus proche de chacun des modes de fonctionnement.

---

[1]OFDM : technique de modulation se basant sur le multiplexage fréquentiel de signaux.

Pour les deux classes de problème présentées (génération de parties calculatoires et d'entrelaceurs), on s'aperçoit que les difficultés se situent au niveau de l'architecture mémoire qui devra assurer le réordonnancement et la temporisation (contraintes spatio-temporelles) des données, tout en minimisant les coûts en terme de surface et de consommation. Malheureusement, les outils de synthèse de haut niveau actuels ne sont pas adaptés pour générer ce type de fonction, notamment concernant la prise en compte au sein d'une même architecture de différents modes de fonctionnement de façon optimisée.

### 1.2 Etat de l'art

Parmis les travaux concernant la synthèse d'interfaces, on peut citer ceux de [Hommais, 2001] qui, en se basant sur des modèles (*template*) de communication, présente une architecture générique pour la synthèse d'interfaces (approche *plateform-based*). Dans [Abbes, 2004], les auteurs utilisent une architecture basée sur des multiplexeurs/démultiplexeurs et des FIFO. Dans ces deux approches, les auteurs supposent que les séquences de données produites sont identiques aux séquences de données consommées (pas d'adaptation sur les ordres de production/consommation des données). De plus, les FIFO sont dimensionnées par simulation (approche *set and simulate*).

Dans [Turjan, 2002] l'approche vise à déterminer, à la compilation, si une FIFO est suffisante pour toute paire producteur/consommateur d'un réseau de processus de Khan. Quand la séquence de données produite est différente de celle des données consommées, il faut alors ajouter une couche de mémorisation et de contrôle spécifique, comme proposé dans [Zissulescu, 2002]. Cette couche supplémentaire inclue une CAM (Content Addressable Memory) dans laquelle les données sont accédées par l'intermédiaire d'une table de hachage. Si cette implémentation permet de gérer séquences d'échanges de données non déterministes (ordre) entre les IP, elle ne permet pas de minimiser l'adaptateur puisque le recouvrement entre les entrées et les sorties n'est pas permis. Dans [Baganne, 1998] une méthode formelle est proposée pour la conception d'interfaces matérielles. Un modèle d'interface générique, ciblé par l'outil de synthèse, est proposé. Les contraintes d'Entrées/Sorties (E/S) de bas niveau peuvent inclure des spécifications temporelles strictes ou des ordres de transfert. La synthèse d'interface est réalisée par une procédure d'allocation de composant mémoire (FIFO, LIFO, Registre) Toutefois, le dimensionnement de ces éléments n'est pas abordé.

Dans [Vermeulen, 2000] les auteurs proposent une méthodologie de réutilisation d'IP pour lesquelles les circuits sont décrits en trois niveaux. Les transferts de données et les optimisations de stockage sont réalisés par réorganisation d'indices et imbrication de boucles. Malheureusement, les auteurs ne présentent pas la méthodologie utilisée pour produire le composant RTL[2] à partir de la spécification algorithmique.

Enfin, dans [Coussy, 2005], les auteurs proposent un ensemble de techniques dédiées à la conception d'applications de type DSP. La synthèse de haut niveau de l'unité de calcul est réalisée sous contrainte d'E/S et d'architecture. L'approche produit un chemin de donnée optimisé mais requière toujours la conception séparée d'une unité de communication.

De notre point de vue, une approche globale pour la conception d'adaptateur spatio-temporel doit résoudre trois classes de problème. L'approche doit fournir (1) une modélisation des contraintes d'E/S, (2) une analyse de celles-ci et (3) des techniques de synthèse de haut niveau pour générer l'architecture RTL du composant. C'est ce que nous proposons de présenter dans cet article. Celui-ci est organisé comme suit : nous présentons d'abord une formulation générale du problème. Puis, nous développons l'approche et la modélisation formelle que nous proposons. Enfin, nous présentons les premiers résultats que nous avons obtenus pour la conception d'un entrelaceur (brassage de données).

### 1.3 Formulation du problème

Considérons un exemple simple : une architecture incluant deux composants qui échangent un ensemble de données $S = \{a, b, c, d, e\}$. $S$ est produite par le bloc #1 et est consommée par le bloc #2 à travers une connexion point à point.

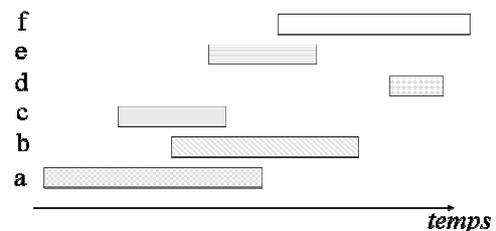

**Figure 1.** Durée de vie des données

Comme on peut le voir sur la Figure 1, l'ordre de production des données est $S_p = (a,c,b,e,f,d)$, c'est à dire $t^p_a < t^p_c < t^p_b < t^p_e < t^p_f < t^p_d$ alors que l'ordre de consommation de ces données est différent, $S_c = (c,a,e,b,d,f)$ c'est à dire $t^c_c < t^c_a < t^c_e < t^c_b < t^c_d < t^c_f$ Cette différence entre les deux séquences E/S peut (1) provenir de l'intégration de deux IP n'ayant pas étés conçus dés le départ pour être assemblés ensemble ou (2) être volontairement recherché dans l'algorithme, par exemple pour introduire de la redondance cyclique (pas dans cet exemple). Il est alors nécessaire d'introduire un adaptateur spatio-temporel entre les blocs #1 et #2. Ce composant supplémentaire peut être conçut en utilisant une mémoire ou une *mer* de registres (6 dans l'exemple de la Figure 1). Toutefois, ces deux premières solutions d'implémentation peuvent être améliorées en terme de surface et de performances en concevant un composant optimisé incluant des éléments mémorisant spécifiques. Ce composant, que nous appellerons **STAR** pour *Space-Time AdapteR*, se compose de chemins de données et de machines d'états associées pour implémenter le contrôle (cf. Figure 2).

---

[2] RTL : se dit d'une modélisation synthétisable (Register Transfert Level)

*Définition 1:* un **adaptateur spatio-temporel (STAR)** est un composant qui réorganise les séquences de données dans une approche déterministe, pour un mode de fonctionnement donné.

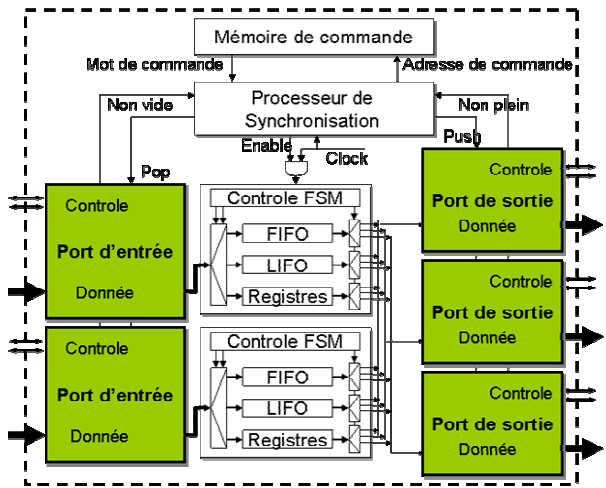

**Figure 2.** Architecture type d'un composant STAR

Le STAR fonctionne selon une architecture globalement asynchrone et localement synchrone, basée sur le LIS décrit dans [Bomel, 2005]. La logique d'interconnexion gère la répartition des données depuis leurs ports d'entrée vers les éléments mémorisant, puis de ces derniers vers leurs ports de sortie. Les automates de contrôle gèrent les protocoles de communication de chaque port et génèrent les signaux de contrôle pour les composants du chemin de données.

Les données arrivant sur les différents ports d'entrées sont mémorisées dans un chemin de donnée composé de FIFO, LIFO et/ou Registre, qui implémentent le stockage et le réordonnancement des données. Ce réordonnancement se fait dans l'espace (une donnée entrant par un port d'entrée lambda peut être lue sur n'importe quel port de sortie) et dans le temps (la donnée est stockée pour toute sa durée de vie).

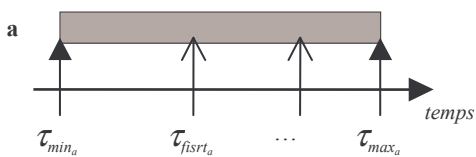

**Figure 3.** Durée de vie de la donnée *a*

*Définition 2:* La **durée de vie** dans un STAR d'une donnée *a* est définie par $\Gamma(a) = [\tau_{min}(a), \tau_{max}(a)]$ où $\tau_{min}(a)$ et $\tau_{max}(a)$ représentent respectivement les dates de *production* et de dernière consommation de *a* dans le composant STAR. La première date de consommation de *a* sera notée $\tau_{first}(a)$ telle que $\tau_{min}(a) < \tau_{first}(a) \leq \tau_{max}(a)$ (cf. Figure 3). Si la donnée n'est lue qu'une seule fois, $\tau_{first}(a) = \tau_{max}(a)$.

Pour générer un STAR, notre approche se base sur la modélisation et l'analyse d'un ensemble de contraintes d'E/S, comme nous allons le voir.

## 2 SOLUTION PROPOSEE

### 2.1 Principe général de l'approche

Nous l'avons déjà indiqué, le point d'entrée de notre approche est une architecture de communication associée à un jeu de contraintes d'E/S: nombre et type de ports, nombre et type de données, relations entre les données et les ports, et enfin les dates d'écriture et de lectures des données. Ces contraintes pour les interfaces du composant STAR sont spécifiées en utilisant la modélisation formelle présentée dans [Coussy, 2005].

La première étape de notre flot, Figure 4, consiste à générer un **G**raphe de **C**ompatibilité des **T**ransferts ou GCT. Ce modèle formel permet l'assignation d'un ensemble de donnée vers un élément de mémorisation unique (FIFO, LIFO, Registre) ainsi que le dimensionnement de ce dernier. Ce premier chemin de données peut ensuite être optimisé en fusionnant les éléments de mémorisation compatibles (durées de vie non recouvrantes). Enfin, la dernière étape consiste à générer la description VHDL de l'architecture qui sera ensuite utilisée pour la synthèse logique.

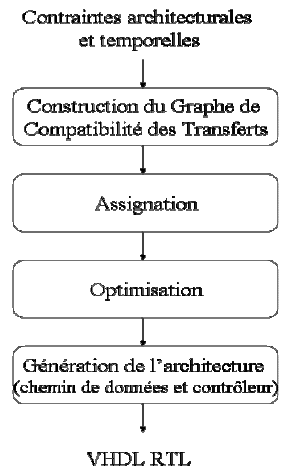

**Figure 4.** Flot de conception

La première étape de notre approche consiste donc à modéliser les contraintes d'E/S sous la forme d'un graphe (GCT). Le jeu de contraintes que nous utilisons contient toutes les informations sur les transferts de données, dont l'ensemble des informations permettant de déterminer les durées de vie de chacune des données transitant dans l'interface (STAR). L'étape suivante consiste à construire un graphe synthétisant l'ensemble de ces informations afin de pouvoir les exploiter.

### 2.1.1 Identification des compatibilités

Les informations modélisées par ce graphe doivent nous permettre de déterminer le type d'élément mémorisant que nous allons utiliser pour générer le chemin de données (FIFO, LIFO ou Registre).

Dans une optique de simplification, nous ne considérons pour l'instant que les contraintes relatives à deux données distinctes. Pour mémoriser ces données, plusieurs solutions s'offrent à nous en fonction des contraintes qui leurs sont associées.

Nous avons développé une fonction d'allocation formelle, **SMA** (**S**torage **M**odule **A**llocation, $\Phi_S$), basée sur les propriétés fonctionnelles de chaque module mémoire (FIFO, LIFO ou Registre). Chaque fonction $\Phi_S$ tient compte des conditions qui doivent être remplies (cf. Table 1) pour chaque paire de données à allouer au même élément mémorisant $\Phi$. Par exemple, pour que deux données puissent être mémorisées dans le même Registre, elles doivent avoir des intervalles de durée de vie disjoints.

- Soit $\Phi$ un élément mémorisant, ($\Phi$ peut être un Registre, une FIFO ou une LIFO),
- Soit $a$ une donnée appartenant à la séquence d'E/S de l'IP,
- Soit $S$ une séquence de données d'E/S partageant la même ressource $\Phi$,

*Définition 3:* la fonction **Storage Module Alloation** (**SMA**) est définie comme une fonction booléenne $\Phi_S$ : $S \rightarrow \{0, 1\}$ où $\Phi_S(a)=1$ si la donnée $a$ peut partager la ressource $\Phi$ avec toutes les données de la séquence $S$.

| Composant mémoire $\Phi$ | Conditions | |
|---|---|---|
| | Contraintes temporelles du STAR | Conditions d'accès |
| Registre | . $\Gamma(a) \cap \Gamma(b) = \emptyset$ | . $(\tau_{min_b} \geq \tau_{max_a})$ |
| FIFO | . $\Gamma(a) \cap \Gamma(b) \neq \emptyset$<br>. $\Gamma(a) \not\subset \Gamma(b)$ | . $(\tau_{min_b} > \tau_{min_a})$<br>. $(\tau_{fisrt_b} > \tau_{max_a})$<br>. $(\tau_{min_b} < \tau_{max_a})$ |
| LIFO | . $\Gamma(a) \cap \Gamma(b) \neq \emptyset$<br>. $\Gamma(a) \subset \Gamma(b)$ | . $(\tau_{min_a} < \tau_{min_b})$<br>. $(\tau_{fisrt_a} > \tau_{max_b})$ |

**Table 1:** Conditions temporelles pour la fonction SMA

A l'aide de ces informations, nous pouvons identifier le type d'élément mémorisant dont nous avons besoin pour stocker deux données distinctes, en respectant un certain nombre de règles :
- Soient $a$ et $b$ deux données,

**Règle 1:** *Compatibilité Registre*
Si $(\tau_{min_b} \geq \tau_{max_a})$ alors on créera un arc étiqueté "Registre".
Ici, les intervalles de durée de vie des données sont dits "non recouvrants". En d'autres termes, ces deux données peuvent être stockées dans le même mot mémoire.

**Règle 2:** *Compatibilité FIFO*
Si $[(\tau_{min_b} > \tau_{min_a})$ et $(\tau_{fisrt_b} > \tau_{max_a})$ et $(\tau_{min_b} < \tau_{max_a})]$ alors on créera un arc étiqueté "FIFO".

Dans ce cas, les intervalles de durée de vie des données sont dits "partiellement recouvrants". Cela signifie que ces deux données peuvent être mémorisées dans une même FIFO, en augmentant la taille au besoin (+1 case mémoire). Nota: la dernière relation ($\tau_{min_b} < \tau_{max_a}$) permet une distinction formelle avec une compatibilité Registre.

**Règle 3:** *Compatibilité LIFO*
Si $[(\tau_{min_b} > \tau_{min_a})$ et $(\tau_{fisrt_a} > \tau_{max_b})]$ alors on créera un arc étiqueté "LIFO".
Dans ce cas, les intervalles de durée de vie des données sont dits "recouvrants-incluants". Nota: la profondeur d'une LIFO est égale au nombre maximum d'éléments qui y sont mémorisés.

**Règle 4:** *pas de compatibilité.*
Dans ce cas, les données sont dites incompatibles. De fait, dans cette situation il n'y a pas d'autre possibilité que d'utiliser deux éléments mémorisant distincts.

*Définition 4:* Un **type de compatibilité** entre deux données est le type d'élément mémorisant qui peut être utilisé pour gérer ces deux données. Cette compatibilité est calculée à partir du jeu de contraintes d'E/S à l'aide de la fonction SMA. Elle se représentera sur le graphe par une étiquette sur l'arc entre ces données (F, L or R respectivement compatibilité FIFO, LIFO ou Registre).

Nous savons maintenant comment identifier le type de compatibilité entre deux données et comment déterminer l'étiquette qui sera sur l'arc reliant ces données s'il existe (cf. section suivante). En étendant cette analyse à l'ensemble des données du fichier de contraintes nous construisons notre graphe.

### 2.1.2 Création du graphe

Nous allons alors construire un graphe de contraintes polaire orienté acyclique G(V,E). L'ensemble des noeuds $V=\{v_0, ..., v_n\}$ représente les données, $v_0$ et $v_n$ étant respectivement le nœud source et le nœud puits. L'ensemble des arcs $E=\{(v_i, v_j)\}$ représente les types de compatibilité entre les noeuds et une étiquette $t_{ij} \in T$ est associée chaque arc $(v_i,v_j)$, $T=\{R,F,L\}$.

*Exemple 1:* Un arc $e_{ij}=(v_i,v_j)$ étiqueté $t_{ij}= L$ signifie que les données ($i$ et $j$) peuvent être mémorisées dans la même structure, si celle-ci respecte une sémantique LIFO.

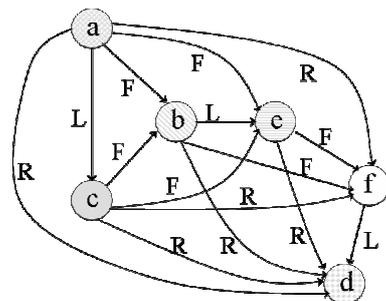

**Figure 5.** Exemple de graphe (contraintes Figure 1)

Pour simplifier les figures, nous ne représenterons pas les noeuds source et puits dans les schémas que nous présenterons.

La construction du graphe implique la création d'arcs entre les données, en respectant l'ordre chronologique de leur arrivée dans le composant STAR. Ainsi, s'il y a $n$ noeuds à instancier, dans le pire cas il y aura $n-1$ arrêtes partant des nœuds directement successeur du nœud source et allant vers tous les autres noeuds, $n-2$ arrêtes partant des nœuds suivants vers tous les nœuds restant (dans l'ordre chronologique) ... Ainsi, dans le pire cas, le graphe contiendra : $n(n-1)/2$ arcs, $O(n^2)$.

## 2.2 Exploitation du graphe

Nous disposons maintenant d'un graphe contenant les informations dont nous avons besoin pour générer une architecture. Pour ce faire la première étape consiste à identifier les différents éléments mémorisant sur le graphe.

### 2.2.1 Identification de structures

Pour identifier les éléments mémorisant sur son modèle de graphe [Baganne, 1998] recherche des cliques de compatibilité puis il en recherche le type, FIFO ou LIFO. On peut néanmoins avoir deux objections à ces travaux (1) la recherche de clique dans un graphe non orienté est un problème NP-complet, (2) l'approche et la modélisation retenues ne permettent pas de dimensionner les éléments identifiés.

Nous avons vu que dans notre modèle de graphe l'information sur le type de structure pour mémoriser deux données est portée par les étiquettes des arcs. L'identification des éléments mémorisant (FIFO ou LIFO) est ainsi simplifiée.

*Définition 5:* un **chemin** est une séquence d'arcs parcourus dans la même direction. Pour qu'un chemin existe entre deux noeuds, il faut qu'il soit possible de se déplacer entre ces noeuds par une séquence d'arcs ininterrompue.

Comme nous allons le voir par la suite, le fait de parcourir un chemin d'un type donné (FIFO ou LIFO) correspond par construction à une clique de compatibilité tel que proposé dans [Baganne, 1998].

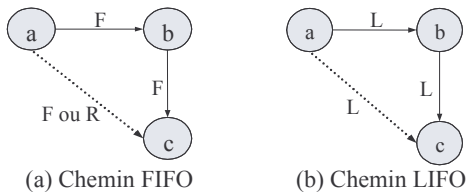

(a) Chemin FIFO      (b) Chemin LIFO
**Figure 6:** Cliques de compatibilité

*Théorème 1*
Soient $a$, $b$, $c$ trois données distinctes en ordre chronologique ($\tau_{min_a} < \tau_{min_b} < \tau_{min_c}$),
**Si $a$ est compatible FIFO avec $b$ et que $b$ est compatible FIFO avec $c$,**
    **Alors, $a$ est compatible FIFO (ou Registre) avec $c$ par transitivité.**

*Preuve :*
Comme nous l'avons vu précédemment, la relation FIFO entre $a$ et $b$ se traduit ainsi: $\tau_{min_b} > \tau_{min_a}$   (1)
$$\tau_{first_b} > \tau_{max_a} \quad (2)$$
$$\tau_{min_b} < \tau_{max_a} \quad (3)$$
De la même façon, la relation FIFO entre $b$ et $c$ peut être traduite ainsi: $\tau_{min_c} > \tau_{min_b}$   (1')
$$\tau_{first_c} > \tau_{max_b} \quad (2')$$
$$\tau_{min_c} < \tau_{max_b} \quad (3')$$
Par transitivité de la relation d'ordre on peut donc écrire :

$\tau_{min_b} > \tau_{min_a}$ et $\tau_{min_c} > \tau_{min_b}$ => $\tau_{min_c} > \tau_{min_a}$
$\tau_{first_c} > \tau_{max_b}$ et $\tau_{first_b} > \tau_{max_a}$ et $\tau_{max_b} \geq \tau_{first_b}$ => $\tau_{first_c} > \tau_{max_a}$

Nous obtenons: $\tau_{min_c} > \tau_{min_a}$
$\tau_{first_c} > \tau_{max_a}$

Ainsi, les données $a$ et $c$ ne peuvent pas être compatible LIFO puisque $\tau_{first_c} > \tau_{max_a}$, mais elles sont bien compatibles FIFO par définition. Toutefois, les relation (3) et (3') permettant de faire la distinction entre compatibilité FIFO et Registre ne peuvent être retrouvées dans le résultat final. Puisque ces relations sont utilisées pour distinguer les compatibilités F et R nous n'avons pas suffisamment d'information pour pouvoir faire cette distinction sur l'arc induit. Donc la compatibilité entre $a$ et $c$ peut être FIFO ou Registre (cf. Figure 6.a).

*Lemme 1:*
Un chemin de données compatibles FIFO, $P_F$, est par construction une clique de compatibilité regroupant un ensemble de données pouvant être mémorisées dans une même FIFO.
Ceci peut être prouvé en appliquant récursivement le *Théorème 1* à l'ensemble des noeuds du chemin $P_F$.

Nota: comme on ne peut déterminer si la relation induite peut être FIFO ou Registre, la profondeur de la FIFO n'est pas égale au nombre d'éléments du chemin (cf. *Règle 1*). Nous aurons besoin d'une approche particulière pour dimensionner une FIFO, ce que nous allons voir par la suite.

*Théorème 2*
Soient $a$, $b$, $c$ trois données distinctes en ordre chronologique ($\tau_{min_a} < \tau_{min_b} < \tau_{min_c}$),
**Si $a$ est compatible LIFO avec $b$ et que $b$ est compatible LIFO avec $c$,**
    **Alors, $a$ est compatible LIFO avec $c$ par transitivité.**

*Preuve :*
Comme nous l'avons vu précédemment, la relation LIFO entre $a$ et $b$ se traduit ainsi: $\tau_{min_b} > \tau_{min_a}$
$$\tau_{fisrt_a} > \tau_{max_b}$$
De la même façon, la relation LIFO entre $b$ et $c$ peut être traduite ainsi: $\tau_{min_c} > \tau_{min_b}$
$$\tau_{fisrt_b} > \tau_{max_c}$$

Par transitivité de la relation d'ordre on peut donc écrire :

$\tau_{min_b} > \tau_{min_a}$ et $\tau_{min_c} > \tau_{min_b} \Rightarrow \tau_{min_c} > \tau_{min_a}$

$\tau_{fisrt_a} > \tau_{max_b}$ et $\tau_{fisrt_b} > \tau_{max_c}$ et $\tau_{max_b} \geq \tau_{first_b} \Rightarrow \tau_{fisrt_a} > \tau_{max_c}$

Nous obtenons: $\tau_{min_c} > \tau_{min_a}$

$\tau_{fisrt_a} > \tau_{max_c}$

Donc les données $a$ et $c$ sont compatibles LIFO par définition (cf. Figure 6.b).

***Lemme 2:***
Un chemin de données compatibles LIFO, $P_L$, est par construction une clique de compatibilité regroupant un ensemble de données pouvant être mémorisées dans une même LIFO.
Ceci peut être prouvé en appliquant récursivement le *Théorème 2* à l'ensemble des noeuds du chemin $P_L$.

Pour la suite de cet article, nous supposerons que les données ne sont accédées en lecture qu'une seule fois (c'est à dire, $\tau_{max_a} = \tau_{first_a}$). Ceci simplifiera la représentation des exemples à suivre, sans influer sur la validité des formules proposées.
Une fois les différents éléments mémorisants identifiés il faut les dimensionner, c'est que nous allons voir.

### 2.2.2 Dimensionnement des structures

La profondeur d'une LIFO est égale au nombre maximum d'éléments que celle-ci peut stocker. Nous allons donc chercher le chemin de compatibilité LIFO $P_L$ le plus grand. Grâce au *lemme 2*, nous savons que les noeuds du chemin $P_L$ constituent une clique de compatibilité LIFO. Puisque ce chemin est le plus grand, il contient donc autant de noeud que de données à mémoriser dans la LIFO (cf. Figure 7).

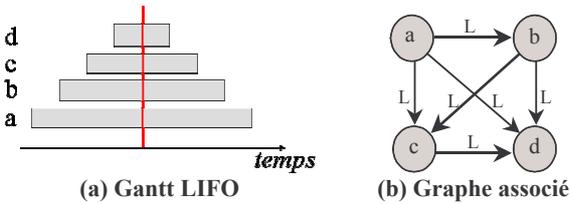

**(a) Gantt LIFO**   **(b) Graphe associé**
**Figure 7:** Exemple d'une LIFO de profondeur 4

Concernant les FIFO, le *lemme 1* nous indique que toutes les données d'un chemin de compatibilité FIFO peuvent être mémorisées dans une seule et même FIFO, puisqu'elles constituent une clique de compatibilité.
Toutefois, nous avons remarqué que nous ne pouvions déterminer par simple parcours du chemin si les arcs induits portaient des compatibilités FIFO ou Registre. La profondeur d'une FIFO n'est donc pas égale au nombre maximum d'éléments qu'elle peut mémoriser, comme le montre l'exemple de la Figure 8. Notre modélisation permet de proposer une solution simple : si on observe le Gantt des contraintes de la Figure 8.a on se rend compte que la profondeur de la FIFO est égale au nombre maximum de données toutes recouvrantes entre elles parmi toutes les données du chemin. Nous pouvons alors proposer une règle simple pour calculer la profondeur de la FIFO :

***Théorème 3***
Soit $P$ le plus grand chemin de compatibilité FIFO,
Soit $i$ un noeud du graphe appartenant à $P$,
Soit $S_i$ le nombre d'arc étiqueté FIFO entrant dans le noeud $i$ et provenant d'un autre noeud du chemin $P$,
Alors,
   **Profondeur = 1 + max ({ $S_i$ | pour tout noeud $i$ de $P$}).**

En effet, le problème se ramène à identifier le nombre maximum de données recouvrantes (en fonction des contraintes E/S) dans le chemin $P$. De par la procédure de construction du graphe que nous avons présenté, pour un noeud donné cela revient à dénombrer les arcs entrants étiquetés "F" en provenance du chemin $P$.

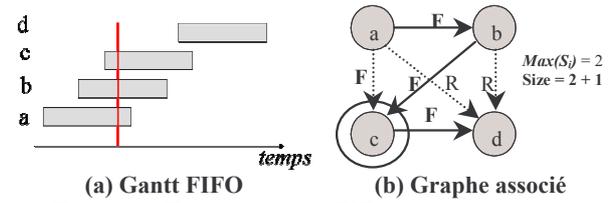

**(a) Gantt FIFO**   **(b) Graphe associé**
**Figure 8:** Exemple d'une FIFO de profondeur 3

L'algorithme correspondant est présenté Figure 9. On remarquera que celui-ci tire parti (1) de l'orientation de notre modèle de graphe (commencer par le dernier noeud du chemin) pour optimiser le nombre de calcul à faire, et (2) de la construction des arcs dans l'ordre d'arrivé des données, en stoppant dès qu'il n'est plus possible de trouver une profondeur plus importante (c'est à dire le nombre de noeud non traités est inférieur à la profondeur max déjà calculée).

---

$P_{FIFO}$ ; *// chemin FIFO*
$Nb_{cur}$; *// nombre d'arcs FIFO entrant dans le noeud courant, en provenance de $P_{FIFO}$*
FIFO_Size = 0; *// entier*

Partir du dernier noeud de $P_{FIFO}$;
**Tant que** (il reste suffisamment de noeuds dans $P_{FIFO}$ pour obtenir une profondeur plus grande)
**boucle**
  **Si** ($Nb_{cur} >$ FIFO_Size) **alors**
     FIFO_Size = $Nb_{cur}$;
  **FinSi**;
**Fin boucle**;

---

**Figure 9:** Algorithme de dimensionnement FIFO

Dans le pire cas, tous les noeuds du chemin devront être testés (sauf le tout premier), c'est le cas si la profondeur de la plus grande FIFO (en fonction des contraintes d'E/S) est de *2* cases mémoires. Ainsi, pour un chemin donné $P = \{v_0, ..., v_n\}$, le nombre de passage dans la boucle sera en $O(n)$.

### 2.2.3 Assignation

Nous savons maintenant construire d'un modèle de graphe permettant de représenter les informations contenues dans les fichiers de contraintes. Grâce aux propriétés exprimées dans les *lemmes 1* et *2*, nous savons comment identifier sur ce graphe des sous-ensembles de données que l'on peut affecter à un même élément mémorisant. Enfin nous savons comment dimensionner ces structures. Il reste maintenant à déterminer comment nous allons construire une architecture à partir de ce graphe. Sur la Figure 4, cette nouvelle étape est celle d'*assignation* des structures.

Cette étape va consister à identifier dans le graphe toutes les structures FIFO ou LIFO possibles. Nous procèderons en trois étapes (1) identification de la meilleure structure, en fonction d'une heuristique que nous présenterons par la suite, (2) fusion de ces données dans un nœud hiérarchique et enfin (3) identification des structures sur les nœuds non hiérarchiques restant dans le graphe.

On parle ici de nœud hiérarchique car la fusion de plusieurs nœuds-données dans un tel nœud demande d'y adjoindre une information supplémentaire pour l'étape suivante d'*optimisation* (cf. Figure 4) : la durée de vie de cette structure, c'est à dire le lapse de temps pendant lequel elle est utilisée. Celui-ci se construit ainsi :

Soit un chemin de données compatibles, $P=(v_0, ..., v_n)$,
- si P est un chemin de compatibilité FIFO, alors la durée de vie du nœud hiérarchique qui sera créé sera $[\tau_{min_{v_0}}, \tau_{max_{v_n}}]$,
- si P est un chemin de compatibilité LIFO, alors la durée de vie du nœud hiérarchique qui sera créé sera $[\tau_{min_{v_0}}, \tau_{max_{v_0}}]$.

Naturellement, le choix des nœuds fusionnés dans un nœud hiérarchique va influencer l'architecture finale obtenue puisque ces nœuds ne serons plus pris en compte par la suite pour l'assignation des autres structures.

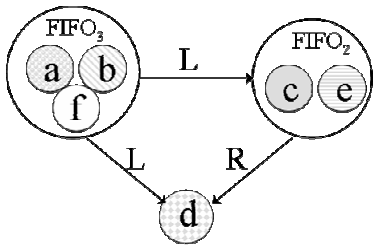

**Figure 10:** Assignation possible du graphe Figure 5.

Pour s'assurer que les choix faits à cette étape ne vont pas dégrader irrémédiablement l'architecture que nous allons obtenir, nous avons dans un premier temps basé cette assignation sur une heuristique utilisant plusieurs métriques, dont l'utilisateur peut pondérer les importances relatives. Ces métriques concernent entre autres la profondeur des structures identifiées, la complexité du démultiplexage associé à une structure ou le taux d'utilisation de la structure.

### 2.2.4 Optimisation

Lorsque l'on ne peut plus trouver de nouvelle structure FIFO ou LIFO sur le graphe, la troisième étape du flot présenté Figure 4 entre en jeu. Nous allons maintenant chercher à optimiser cette première architecture en fusionnant entre-elles les différentes structures, lorsque cela sera possible. Pour ce faire nous allons simplement construire un nouveau graphe, en appliquant la même approche que précédemment, mais en utilisant cette fois les informations sur les durées de vie des structures sauvegardées dans les nœuds hiérarchiques. La différence est que nous allons chercher à fusionner non plus des données, mais des éléments complexes. Donc, pour éviter tout conflit, il ne faut rechercher que des chemins de compatibilité Registre, c'est à dire des structures n'ayant aucun recouvrement temporel de leur durée de vie. Cette fois, lors de la fusion un nouveau type de nœud hiérarchique est créé : il permet de mémoriser plusieurs intervalles de durée de vie, ceux des structures fusionnées dans ce nœud. Ceci permet de garder la possibilité de fusionner ce nœud avec un autre dont les intervalles de durées de vie seraient tous non recouvrant avec ceux de ce nœud.

Ici encore, nous allons utiliser une heuristique basée sur un jeu de métriques de même type que dans l'étape d'assignation, afin de faire les choix de fusion les meilleurs possibles. Lorsque toute les fusions possibles ont étés réalisées, le graphe résultant sera alors utilisé pour générer l'architecture VHDL correspondante.

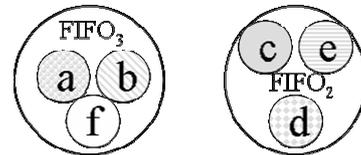

**Figure 11:** Optimisation du graphe de la Figure 10

La Figure 11 montre un résultat possible, après optimisation, pour le jeu de contraintes présenté par la Figure 1. On obtient ici une architecture comportant une FIFO de profondeur 3 et une FIFO de profondeur 2 qui manipule 3 données, pour un total de 6 données à gérer (Gain de 1 case mémoire ici).

### 2.3 Résultats

Une première version d'un outil associé au flot de conception proposé a été développée et testée sur un exemple industriel d'entrelaceur. Ce composant doit être capable de basculer à la volée entre différents modes de fonctionnement. Suivant le mode retenu, le composant doit entrelacer (1) 300, (2) 600 ou (3) 1200 données (sur 1 bit), en respectant des contraintes de débit et de latence données. Dans chaque cas le schéma d'entrelacement est différent. Les résultats des synthèses réalisées par les concepteurs, avec les outils du marché montrent qu'aucun n'est capable de générer une architecture intégrant de façon optimale les différentes configurations du circuit. Dans le meilleur cas, pour la configuration 3, une mer de bascule flip-flop (1200) est générée. De plus, comme le composant est sensé fonctionner en continu, cette architecture de 1200 éléments est doublée et fonctionne en mode ping-pong (un bloc réceptionne une trame de 1200 données pendant que l'autre transmet la trame précédente entrelacée).

Nous avons dans un premier temps produit un outil générant automatiquement un jeu de contraintes en fonction de données fournies par le concepteur : nombre de données, débit, latence, schéma d'entrelacement. Nous avons ensuite utilisé le jeu de contraintes produit pour générer 3 entrelaceurs (un par mode de fonctionnement).

| Mode | Tps CPU (sec.) | RAM (Mo) | Nombre de structures | | | | Plus grande FIFO | Plus petite FIFO | Plus grande LIFO | Plus petite LIFO |
|---|---|---|---|---|---|---|---|---|---|---|
| | | | *FIFO* | *LIFO* | *Reg* | *Total* | | | | |
| 300 | 19 | 13 | 17 | 15 | 0 | 32 | 16 | 2 | 13 | 2 |
| 600 | 191 | 70 | 24 | 17 | 1 | 43 | 25 | 2 | 23 | 4 |
| 1200 | 2058 | 154 | 36 | 17 | 1 | 54 | 50 | 2 | 18 | 6 |

**Table 2 :** Résultats obtenus avec la version actuelle du logiciel

La machine utilisée est un Pentium4 2.4GHz, 512Mo de RAM. Le tableau 2 donne les résultats obtenus : si l'on regarde par exemple le mode 1200 données, on s'aperçoit que l'outil génère une architecture comportant 54 structures différentes (soit 54 éléments différents à piloter alors que la solution de référence suppose d'en piloter 1200 (*2 pour le mode pipeline)). Le gain potentiel en terme de contrôle est donc important. Toutefois, le schéma d'entrelacement est tel qu'il ne permet pas d'optimiser l'architecture en terme de nombre de points mémoire (phase d'optimisation), il en faudra toujours 1200. Le mode pipeline n'est pour l'instant pas pris en compte dans notre outil, mais sur ce point, il apparaît clair que nous pouvons obtenir un nombre de point mémoire significativement plus faible que les 2400 utilisés pour l'instant par la solution de référence.

Les temps de calcul et les besoins en RAM de notre outil sont donnés à titre indicatif.

## 3 CONCLUSION ET PERSPECTIVES

### 3.1 Conclusion

Nous avons présenté un flot de conception permettant de modéliser et de générer une architecture respectant un ensemble de contraintes d'E/S. L'approche repose sur un modèle de graphe original permettant d'identifier et de dimensionner, avec une complexité raisonnable, les éléments mémorisant nécessaires et sur l'architecture de communication associée.

Les premiers résultats sont prometteurs. Pour pouvoir comparer et quantifier précisément les gains obtenus avec notre méthodologie, il faut générer l'architecture en VHDL, ces travaux sont actuellement en cours.

### 3.2 Travaux à venir

La première étape va consister à implémenter la partie génération de code VHDL RTL afin de pouvoir quantifier précisément les gains obtenus.

Il nous faudra également intégrer la gestion du mode pipeline dans notre approche (repliement du graphe de contraintes).

La prise en compte des différentes configurations d'un composant constituera l'étape suivante. Pour cela plusieurs voies s'offrent à nous : identification des compatibilités entre les différents graphes de structures, identification du sous graphe commun aux graphes correspondant aux configurations possibles du composant (plus prometteur, mais plus complexe), …

Enfin, nous devons améliorer de l'exploration de l'espace des solutions avec une approche plus solide encore que nos heuristiques. Des solutions sont à l'étude concernant l'utilisation d'approche de type colonies de fourmies [Blum, 2005] ou génération de colonnes [Briant, 2004].

## BIBLIOGRAPHIE